\documentclass[amsmath]{revtex4}

\usepackage{graphicx}
\usepackage{dcolumn}
\usepackage{bm}

\newcommand{\ket}[1]{|#1\rangle}

\newcommand{\0}{|0\rangle}
\newcommand{\1}{|1\rangle}

\begin{document}

\title{Quantum Multiplexing with the Orbital Angular Momentum of light.}
\author{Juan Carlos Garc\'ia-Escart\'in}
\email{juagar@tel.uva.es}
\author{Pedro Chamorro-Posada}
\affiliation{Departamento de Teor\'ia de la Se\~{n}al y Comunicaciones e Ingenier\'ia Telem\'atica. Universidad de Valladolid.\\ETSI Telecomunicaci\'on. Camino del Cementerio s/n. 47011 Valladolid, Spain.}
\date{\today}
\begin{abstract}
The orbital angular momentum, OAM, of photons offers a suitable support to carry the quantum data of multiple users. We present two novel optical setups that send the information of $n$ quantum communication parties through the same free-space optical link. Those qubits can be sent simultaneously and share path, wavelength and polarization without interference, increasing the communication capacity of the system. The first solution, a qubit combiner, merges $n$ channels into the same link, which transmits $n$ independent photons. The second solution, the OAM multiplexer, uses CNOT gates to transfer the information of $n$ optical channels to a single photon. Additional applications of the multiplexer circuits, such as quantum arithmetic, as well as connections to OAM sorting are discussed.
\end{abstract}
\maketitle

\section{Introduction}
Quantum communication is the most developed area of quantum information. There are a growing number of successful optical implementations of quantum communication protocols like teleportation or quantum cryptography that have been conducted on a variety of real-world situations inside optical fibre networks and through satellite free-space optical links \cite{PFU04,LHB07,VJT08}. 

At this point, quantum networks must face the question of multiple access. The multiple access problem appears when the limited resources of a communication system have to be shared by a certain number of users. Multiple access techniques are essential in classical communications \cite{Skl83,Skl01}. In order to share the channel capacity, the signals of different users are designed to be orthogonal to each other in, at least, one domain. For instance, in optical fibre communications, one popular multiple access technique is Wavelength Division Multiple Access, WDMA, where different users transmit at different wavelengths so that their data can be sent independently over the same physical link. Similarly, in radiofrequency communications, each user is assigned a part of the spectrum in what is known as Frequency Division Multiple Access, FDMA.

There are already some results that extend WDMA and FDMA for quantum key distribution networks over optical fibre \cite{BBG03} and with radiofrequency qubits \cite{OC06}. In this paper, we show how a new resource, the orbital angular momentum of photons, OAM, can be exploited in a novel quantum multiple access technique. Transmission using OAM encoded data has already been proposed and experimentally demonstrated \cite{GCP04}. In that scheme, the data bits are grouped into multivalued symbols that correspond to different OAM states of light. Those techniques can be extended so that many independent communication parties can send their quantum data together. We will propose two models. In the first one, photons with different OAM are sent side by side in the same channel, in a way similar to WDMA, where photons of different wavelength were transmitted on the same link. The second model employs CNOT gates to encode the data of all the users as a multilevelled symbol, which is then transmitted like in the existing OAM-based communication systems.

Section \ref{OAM} introduces the basic concepts of the angular momentum of light. Section \ref{blocks} gives a list of the building blocks of our setups and discusses different alternatives for implementation with standard optical elements. Section \ref{comb} describes a quantum combiner that can merge $n$ single photon channels into one link. Section \ref{mux} puts forward a model for a quantum multiplexer that transfers the data from $n$ optical qubits into a single photon. Section \ref{apps} reviews some of the possible applications of the OAM multiplexer circuit, in particular its connections with quantum arithmetic circuits and OAM sorters. Finally, Section \ref{outlook} closes the paper with a recapitulation of the results and some comments on the strong and weak points of each OAM multiple access approach. 

\section{Orbital Angular Momentum}
\label{OAM}
Light fields, both classical and quantum, can carry angular momentum. The total angular momentum of light has two components, polarization and orbital angular momentum, or OAM. Polarization can be associated to spin and OAM to the azimuthal phase \cite{ABP03}. Both contributions can be studied independently under the paraxial approximation. The separate analysis and manipulation of the OAM allows for a wide range of classical and quantum applications, such as new free-space communication systems, high precision atomic control and higher dimensional quantum information processing \cite{MTT07}.

Beams with different OAM values are defined by phase terms $e^{i\ell\varphi}$, where $\varphi$ is the azimuthal phase and $\ell$ is the OAM index, usually referred to as the winding number or topological charge. The properties of OAM fields can be taken down to the single photon level and open the door for new single photon quantum states.  

OAM is a particularly promising candidate to implement optical $d$-dimensional quantum information units, the qudits. Photons can be in different orthogonal $\ket{\ell}$ states that carry an $\ell\hbar$ OAM. There is an infinite number of such orthogonal states, one for every different integer value of $\ell$. There also exist fractional OAM states \cite{GOP08}, but they will not be covered here.

Light carrying an $\ell\hbar$ OAM can be readily generated by many alternative techniques. For instance, light with an $e^{i\ell\varphi}$ phase can be created from Hermite-Gauss beams using either spiral phase plates \cite{TRS96}, computer-generated holograms \cite{HMS92} or cylindrical lenses \cite{AV91}. At the quantum level, spontaneous parametric down-conversion, SPDC, provides OAM entangled photon pairs \cite{MVW01}.

OAM photon states can be manipulated with standard optical elements and have been demonstrated to be good carriers of quantum information \cite{MTT01,LPB02}. For all those reasons, they seem to be an ideal embodiment for the ideas of quantum multiplexing. 

\section{Building blocks}
\label{blocks}
The proposed systems will work with single photon generators and dual-rail qubits. We will assume single photons can be produced on demand and that the user information is encoded in the path dual-rail representation, a photon occupying a different position mode for logical $\0$ and logical $\1$. 

The usual notations for photon number, OAM and logic qubit states collide. By default, our quantum states will be OAM states. In order to avoid confusion, we will write logical qubit states with boldface numbers, as $\ket{{\mathbf 0}}$ and $\ket{{\mathbf 1}}$, and the winding number inside OAM states in italics so that $\ket{{\mathit 0}},\ket{{\mathit 1}},\ket{{\mathit 2}}\ldots$  are the states with an OAM of $0, \hbar, 2\hbar\ldots$ All these OAM states are states with a single photon. The vacuum state, of zero photon number, will be represented as $\ket{v}$.

We will present the OAM multiplexing system from a black box approach. First, we will describe the basic blocks that will be needed and then use them as if they were perfect. In this section, we will provide possible physical implementations and discuss limitations and alternatives for each block.

\subsection[Holograms]{Holograms} 
\begin{figure}[h]
\centering
\includegraphics{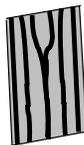}
\caption[Hologram representation.]{Hologram representation. The exact value of $\Delta\ell$ is included below the symbol.\label{hologram}} 
\end{figure}

Special holograms can be designed so that any photon traversing them increases or decreases its winding number, $\ell$, by a fixed amount $\Delta \ell$ of choice \cite{MVW01}. These holograms will be represented by the symbol of Figure \ref{hologram}, which reproduces their characteristic forked interference pattern.

A beam meeting a hologram is divided into components in the different output diffraction orders. The output beam at the $n$th order will suffer an increase of $n\Delta \ell$ in its winding number value. We implicitly assume that only the first diffraction order is selected. $\Delta \ell$ depends on the phase pattern of the hologram. Different patterns, corresponding to different gratings and different $\Delta \ell$ values can be generated with the help of a computer. Computer-generated holograms have been successfully used in many OAM experiments, classical and quantum. 

However, holograms normally introduce important losses, which can be critical at the single photon level. In order to reduce the effect of the losses, the holograms can be built with blazed gratings that maximize the transmitivity in the first order of diffraction \cite{ADA98,VWZ02}. Usually, the higher the value of $\Delta \ell$, the most difficult it is to obtain an efficient hologram. 

Holograms are not the only alternative for producing this increase or decrease of $\ell$. In the following, the word hologram should be understood in a broad sense as the name for any block capable of the transformation
\begin{equation}
\ket{\ell}\stackrel{\Delta \ell}{\longrightarrow}\ket{\ell+\Delta\ell},
\end{equation}
for any input state $\ket{\ell}$. Most of the $\ket{\ell}$ state generation techniques, like the use of spiral phase plates, can be adapted for this purpose. Of particular importance for scalable design is the introduction of spatial light modulators, which can dynamically switch between different hologram patterns and allow an external classical control on the value of $\Delta\ell$ without additional realignment of the optical system \cite{GCP04,YFC06,SGJ07}. There are also liquid crystal based configurable spiral phase plates that can give a similar behaviour \cite{WSS05}.

\subsection[50\% beamsplitters]{50\% beamsplitters and mirrors} 
\begin{figure}[h!]
\centering
\includegraphics{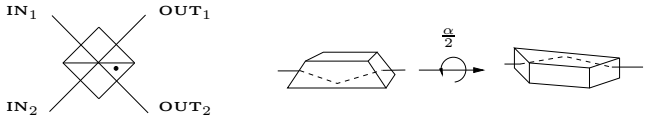}
\caption{Basic optical building blocks. Left: 50\% beamsplitter. Right: Dove prisms.\label{BSDove}} 
\end{figure}

Beamsplitters will be used to divide photons impinging on them into superpositions with one half of the photon's wavepacket on each output port. In the given representation of the 50\% beamsplitter (see Figure \ref{BSDove}, left), the parts of the superpositions resulting from reflection on the side with the dot suffer a sign shift. In our schemes, beamsplitters will always appear as a part of an interferometer. 

We will direct light with the help of mirrors. Mirrors must be used with caution. The phase profile of an $\ket{\ell}$ state is altered on reflection and gives a $\ket{-\ell}$ state. This effect can be easily taken into account in the designs.

\subsection[Dove prisms]{Dove prisms} 
\label{Dove}
Dove prisms rotate any incoming light front by a certain angle \cite{BW97}. Two Dove prisms with a relative angle of $\frac{\alpha}{2}$ (Figure \ref{BSDove}, right) rotate the passing beams through an angle of $\alpha$ with respect to each other. This rotation introduces an OAM dependent phase shift of $\alpha \ell$ between photons that traverse the first and the second prism. When light reaches a Dove prism, it is refracted into a path that reflects from the bottom plane of the prism and then is refracted again out of the prism with the corresponding rotation. Inside the Dove prism, apart from the phase effect, there is a reflection that changes the sign of the winding number. An OAM state that goes through an $\frac{\alpha}{2}$ rotated prism will show the evolution 
\begin{equation}
\ket{\ell}\stackrel{Dove}{\longrightarrow}e^{i\ell\alpha}\ket{-\ell}.
\end{equation}

We will make use of Dove prisms inside interferometric setups. Having two prisms performs the rotation in a symmetric way without compromising path stability by introducing an additional optical path only in one of the arms.  

\subsection{Sorting interferometer}
\begin{figure}[h!]
\centering
\includegraphics{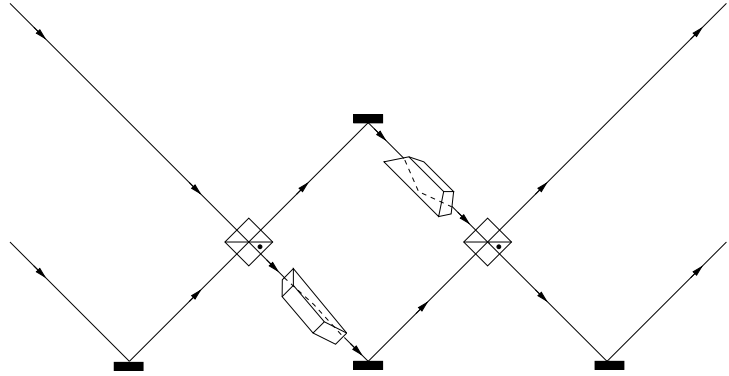}
\caption{W sorting interferometer.\label{Inter}} 
\end{figure}

The basic unit for our proposals will be a sorting interferometer like the OAM sorter blocks of \cite{LPB02}. In it, inputs of different OAM will change their path or not depending on their winding number. Figure \ref{Inter} shows the whole setup, made of two 50\% beamsplitters, four mirrors and two Dove prisms. 

\begin{figure}[h!]
\centering
\includegraphics{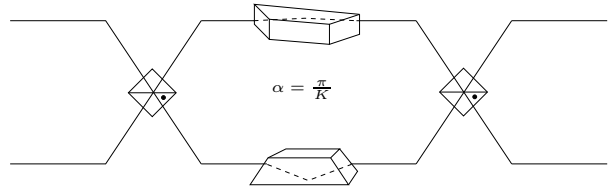}
\caption[Interferometer with two Dove prisms with a relative angle of $\frac{\alpha}{2}$.]{Interferometer with two Dove prisms with a relative angle of $\frac{\alpha}{2}$. The representation will always include the angle $\alpha$ of the relative rotation of the lower arm beam with respect to the upper arm beam.\label{Doveint}} 
\end{figure}

The given W-shaped setup takes into account the total number of reflections so that the photons reach the Dove prisms in the same $\ket{\ell}$ state they entered the interferometer and, at the output, have suffered an even number of sign shifts and $\ell$ suffers no change. This includes reflections on the beamsplitters and inside the Dove prisms. Additionally, the path lengths have been chosen so that all the photons travel the same distance. This configuration guarantees that interference takes place as expected. In our discussion, we will work with a simplified conceptual representation (Figure \ref{Doveint}). In the following, the effects of reflections will be ignored and we will study the sorting interferometer only in terms of the rotation that the Dove prisms introduce. 

We can consider the effect of the sorting block on photons with different OAM, both for inputs in the upper and in the lower faces of the first beamsplitter. All the photons will be in a superposition of OAM states with indices that are multiples of a constant K, which is related to the angle between the Dove prisms so that $\alpha=\frac{\pi}{K}$. Our inputs are, then, superpositions of $\ket{\ell}$ states such that $\ell=mK$ for any integer $m$. For this restriction, the more general photon superposition can be written down as $\ket{\psi}=\sum_m \alpha_m \ket{mK}$.

Imagine $\ket{\psi}$ enters the upper port of the interferometer. After the beamsplitter we have,
\begin{equation}
\sum_m \alpha_m \frac{ \ket{mK}\ket{v}+\ket{v}\ket{mK}}{\sqrt{2}}.
\end{equation}
There is only one photon. The beamsplitter takes this input photon into an equal superposition of being in the upper or the lower arm. For $\alpha=\frac{\pi}{K}$ and our OAM states in which $\ell$ is a multiple of K, the Dove prisms will introduce a phase shift of $\ell \frac{\pi}{K}=m\pi$ in the lower path photons. This produces the state
\begin{equation}
\sum_m \alpha_m \frac{ \ket{mK}\ket{v}+(-1)^m\ket{v}\ket{mK}}{\sqrt{2}}.
\end{equation}
At the second beamsplitter, the states with an even $m$ will have suffered no change and the original state is restored at the upper port. States with an odd $m$, on the other hand, will interfere destructively in the upper port and constructively in the lower port. This system works as a sorter that directs all the even multiples of K to the upper port and the odd ones to the lower. 

The same reasoning can be applied for inputs on the lower port. The evolution can be summed up as
\begin{equation}
\sum_m \alpha_m \ket{v}\ket{mK}\stackrel{BS_1}{\longrightarrow} \sum_m \alpha_m \frac{ \ket{mK}\ket{v}-\ket{v}\ket{mK}}{\sqrt{2}}\stackrel{Dove}{\longrightarrow} \sum_m \alpha_m \frac{ \ket{mK}\ket{v}-(-1)^m\ket{v}\ket{mK}}{\sqrt{2}},
\end{equation}
where both modes, up and down, have been considered from the beginning. In the even $m$ case, the sign shift coming from the reflection at the first beamsplitter will create a constructive interference at the lower port and a destructive interference at the upper one. Again, for an odd $m$, the extra sign shift creates a change in the output port. This kind of operation, conditional on the OAM state, can be useful for later manipulations and even for the construction of different phase gates. 
 
The presented interferometric setup can be subject to imperfections. For instance, Dove prisms can introduce astigmatism for highly focused beams, but this problem can be avoided by a proper design \cite{GMT06}. We will assume, as usual, that the Dove prisms produce a perfect $\alpha\ell$ phase shift. 

There exist alternatives to the use of Dove prisms. Graded-index (GRIN) rods with a quadratic index profile and certain configurations of bulk optical lenses can produce the same $\ell$ dependent phase shift as the Dove prisms and can be used in similar interferometric setups called spatial modal interleavers \cite{XWK01}. The value of the corresponding K factor can be adjusted by varying the rod's length. 

\subsection[Photodetectors]{Photodetectors} 
Although not strictly necessary for the multiplexing, photodetectors can be introduced to check for operation failures. We will use photodetectors to ascertain the absence or presence of a single photon. If everything is working properly, no photon will appear on certain ports. 

Efficient single photon detection can be problematic, even in this case where exact photon counting is not required. Avalanche photodiodes, APDs, are an interesting option for these yes/no tasks of informing if there are any photons or none, but dark counts can induce to error. Nevertheless, photodetectors are not essential to any of the given applications. In our schemes, even small efficiencies will improve the global operation as long as the false counts are minimized. Early detection of errors can save unnecessary further processing. 

At some point of the optical quantum computation, photodetectors might be needed. This is an important source of error for many optical implementations, but it is common to all the proposals. They are not introduced by the multiple access stage and will not be treated here.   
 
\section{Qubit combiner}
\label{comb}
The first multiple access setup will use only off-the-shelf optical components that have already demonstrated good behaviour in previous OAM quantum information experiments. This multiplexer will be called qubit combiner to distinguish it from the more sophisticated alternative model of the next section. 

The OAM combiner will allow to merge N qubits into a single spatial channel by transforming their dual-rail path encoding into OAM dual-rail. The result will be that {\bf$\mathbf N$ qubits} will be transmitted by {\bf$\mathbf n$ photons} in {\bf one channel}.

By encoding the different qubits in states orthogonal to each other, they can be sent together without interferences between them. In many aspects, the scheme works like WDMA, where $n$ photonic channels are sent in parallel through a single fibre, ideally without affecting each other because of their different wavelength modes. 

\subsection{OAM dual-rail conversion}
\begin{figure}[h]
\centering
\includegraphics{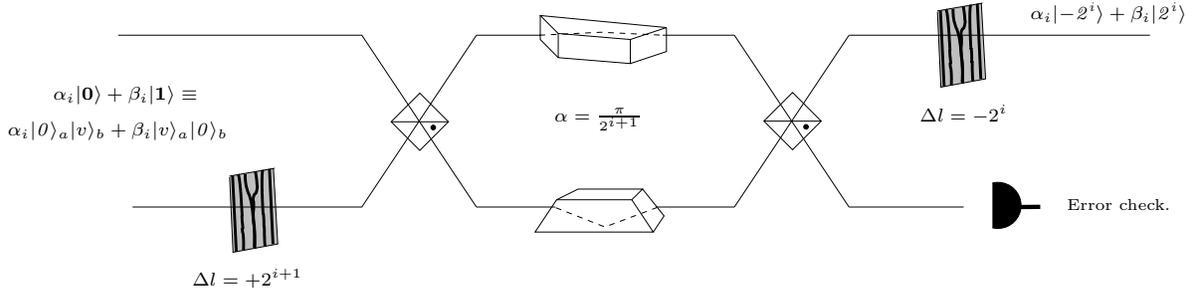}
\caption{Path dual-rail to OAM qubit conversion.\label{converter}} 
\end{figure}

A previous necessary step for the combination is displacing the qubit states to appropriate OAM subspaces. The setup of Figure \ref{converter} takes the path dual-rail qubits of user $i$ to a $\{\ket{\mathit{-2^i}},\ket{\mathit{2^i}}\}$ state space. In that form, qubits can be later merged into the same channel.

The input is supposed to be a path dual-rail qubit whose photon carries no orbital angular momentum. The part of the superposition that represents the logical $\ket{\mathbf{0}}$ will not be affected by the Dove prisms and will go out the interferometer in the upper port. The part that encodes the logical $\ket{\mathbf{1}}$, however, meets a hologram before entering the interferometer that will transform the input of the lower port into $\ket{\mathit{2^{i+1}}}$. This state is divided into both arms so that the part of the photon originally in the lower arm suffers a $\pi$ sign shift as a result of the Dove prisms. This means that both components of the photon will combine in the upper port in the state $\alpha_i\ket{\mathit{0}}+\beta_i\ket{\mathit{2^{i+1}}}$. All the other qubits will follow a similar procedure. In order for the logical $\ket{\mathbf{0}}$ states of different qubits not to interfere, an additional $-2^i$ displacement is introduced so that the qubit is now in an OAM dual-rail encoding $\alpha_i\ket{\mathit{-2^i}}+\beta_i\ket{\mathit{2^{i}}}$. The name dual-rail is still appropriate as the encoding is based on two orthogonal modes. 

This encoding could also come from the original qubits to avoid the presence of lossy holograms, as long as each source is assigned a fixed index $i$ from the beginning. Each user would then work with OAM qubits of the corresponding power of two. OAM qubits could be manipulated in a way much similar to the path or polarization dual-rail qubits. Dove prisms can produce phase shift gates and cylindrical lenses \cite{PA02} or mirrors can convert $\ket{\ell}$ modes into $\ket{-\ell}$ modes providing for a NOT gate in the new encoding. Two qubit gates for this new encoding present the same difficulties as in the path dual-rail case, as there are still two photons that need to interact.

\subsection{OAM Qubit merger}
\begin{figure}[h]
\centering
\includegraphics{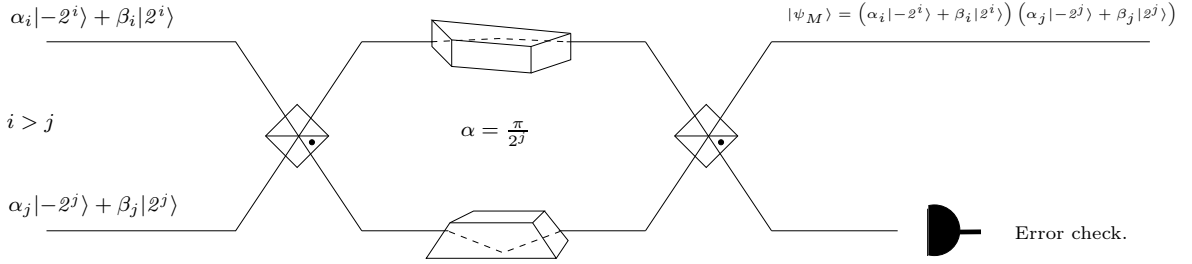}
\caption{OAM qubit merger.\label{merger}} 
\end{figure}

Once the OAM dual-rail qubits are available, they can be put in a superposition in a single path. Qubits will be merged one by one with the help of the optical circuit of Figure \ref{merger}.

Imagine two qubits from sources of indices $i$ and $j$ such that $i>j$. The qubits are encoded into $\ket{\ell}$ states of the powers of two of the source index. For Dove prisms with $\alpha=\frac{\pi}{K}=\frac{\pi}{2^{j}}$, the $j$th qubit will be expressed in an odd, -1 or 1, multiple of the prisms' K, but any higher power of two will always be an even multiple. 

As a result, qubits encoded in the $\{\ket{\mathit{-2^i}},\ket{\mathit{2^i}}\}$ basis will suffer no phase shift and will exit the interferometer in the upper port, while the $j$th qubits will cancel for the lower port as a consequence of the Dove induced phase shift. They will also go out the upper port, which now has two photons that form the multiplexed state
\begin{equation}
\ket{\psi_M}=\alpha_i\alpha_j\ket{\mathit{-2^i}}\ket{\mathit{-2^j}}+\alpha_i\beta_j\ket{\mathit{-2^i}}\ket{\mathit{2^j}}+\beta_i\alpha_j\ket{\mathit{2^i}}\ket{\mathit{-2^j}}+\beta_i\beta_j\ket{\mathit{2^i}}\ket{\mathit{2^j}}.
\end{equation}

The qubit merger will also work for the outputs of the previous stages. As long as a descending OAM order is followed, all the terms in the upper port superposition will behave exactly like in the qubit case. They are even multiples and the new qubit in the lower port is an odd multiple that will suffer a sign shift. 

Repeating the merging for the $n$ qubits, the final product state put on the channel will be the tensor product
\begin{equation}
\overset{n-1}{\underset{i=0}{\otimes}} \left( \alpha_i \ket{\mathit{-2^i}}+\beta_i\ket{\mathit{2^i}}\right).
\end{equation} 
The $n$ photons coexist in orthogonal states without interfering. They can share the same position and wavelength and the users can send their information whenever it is ready, as many photons can occupy the channel at the same time.

All the operations are reversible and the same elements, in reverse order, can be used to recover the original qubits in a demultiplexer. The photodetectors will find no photons in a perfect operation and act only as an error check. Notice that the merging order must be strictly observed. Some qubits can be skipped, but under no circumstances must a lower index qubit be merged before an upper index one. In the proposed setup, phase shifts must be either of $\pi$ or of $0$. This is behind the somewhat wasteful approach of leaving empty valid $\ket{\ell}$ states between the $\ket{\mathit{2^i}}$ OAM states. Values of $\ell$ that are not a power of two cannot be efficiently extracted with this interferometric approach. They would suffer partial phase shifts similar to the ones that appear if combination does not respect the given qubit order. 

The ordering is easy to implement. Different communication users usually join the channel at different points of the link. We can assign the higher indices to the users that are most distant to the end point and the lower indices to the users that join the channel later. With some additional timing considerations, we can even make the qubits of each user to be inserted into the channel at the same time the photons from the previous users reach their point of the link. If the channels are independent, this step can be skipped. The different users can transmit their qubits at any time with the guarantee that no interference with the other qubits will occur.

\section{OAM Multiplexer}
\label{mux}
The OAM multiplexing model can be further refined if additional resources are available. In particular, it would be interesting to be able to send as many channels as possible with the minimum number of photons. The OAM multiplexer (OAM MUX) will transmit {\bf$\mathbf N$ qubits} by {\bf one photon} in {\bf one channel}. 

A reduction in the number of photons, from $N$ to 1, implies that the original qubits must be destroyed. In order to convey the path information to the OAM of the carrier photon and to later erase the resulting correlations, dual-rail CNOT gates are needed. A dual-rail CNOT gate, here, is considered to be any mechanism that switches two photon paths when there is a photon present in a particular path mode. We will assume that the CNOT gate is insensitive to the particular OAM state of the photon. The control can be in any $\ket{\ell}$ state and the OAM of the target photons is preserved.

\subsection{Multiplexing}
In the multiplexing stage we will take the data of the $n$ channels from the computational basis binary states, made out of $n$ qubits, into the OAM state which has an $\ell$ that corresponds to the integer number expressed by the qubits. We will assign each user an index $i$, from $n-1$ to $0$, in decreasing order. The data from the first users will determine the most significant bits of the binary number that will define the final OAM. The multiplexer will transform each $\ket{\mathbf{00}\cdots\mathbf{00}},\ket{\mathbf{00}\cdots\mathbf{01}},\ldots,\ket{\mathbf{11}\cdots\mathbf{11}}$ state into the corresponding OAM $\ket{\mathit{0}},\ket{\mathit{1}},\ldots,\ket{\mathit{2^n-1}}$ state.

In the following, $\ket{\mu_j}$ will represent a multiplexed state that can be in any superposition of $\ket{\ell}$ states for which $\ell$ is a multiple of $2^{j}$. This multiplexed state can carry the data from users $n-1$ to $j$. 

\begin{figure}[h]
\centering
\includegraphics{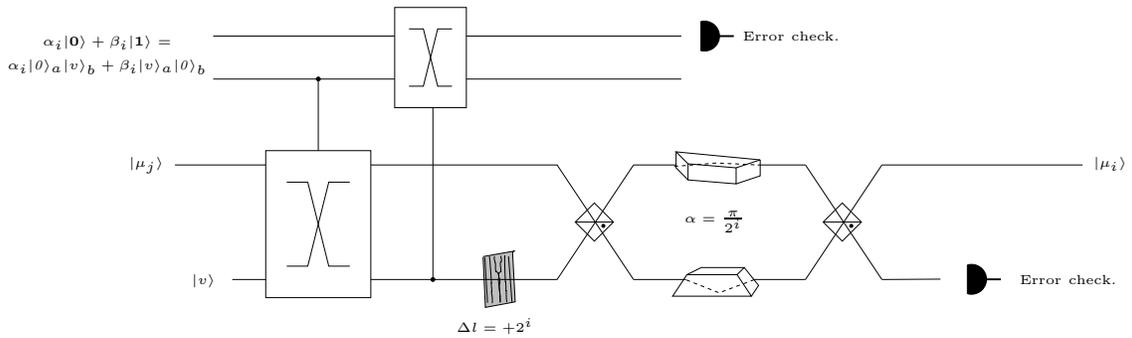}
\caption{OAM multiplexing block for the $i$th qubit.\label{OAMMUX}} 
\end{figure}

Figure \ref{OAMMUX} shows how each qubit is added into the final state. Dual-rail CNOT gates are represented as switches controlled by a photon path. 

For the user of index $i$, the multiplexing block will add an OAM of $2^i$ to all the states of $\ket{\mu_j}$ if the $i$th qubit is $\ket{\mathbf{1}}$. If the $i$th qubit is $\ket{\mathbf{0}}$, the state is not altered. The first dual-rail CNOT gate will convert the incoming qubit-qudit combination $(\alpha_i\ket{\mathbf{0}}+\beta_i\ket{\mathbf{1}})\ket{\mu_{j}}\ket{v}$ into the entangled superposition
\begin{equation}
\alpha_i\ket{\mathbf{0}}\ket{\mu_{j}}\ket{v}+\beta_i\ket{\mathbf{1}}\ket{v}\ket{\mu_{j}}\equiv\alpha_i\ket{\mathit{0}}\ket{v}\ket{\mu_{j}}\ket{v}+\beta_i\ket{v}\ket{\mathit{0}}\ket{v}\ket{\mu_{j}}.
\end{equation}
Then, all the $\ket{\ell}$ states in the lower port go through a $+2^i$ hologram. The states in $\ket{\mu_{j}}$ will all be either the $\ket{\mathit{0}}$ state or the results of previous sums of $+2^j$ for indices $j>i$. Consequently, by themselves, they are not affected by the Dove prisms, which will always induce a multiple of $2\pi$ phase shift. In the given setup, this means that the states on the upper input port of the interferometer, which are associated to $\alpha_i$, will not be affected by it and will go out also on the upper port. After the $+2^i$ sum, the part of the superposition that took the hologram path and is associated to $\beta_i$ will suffer and odd number of $\pi$ phase shifts. The resulting sign change will guide them to the upper port where they will join the rest of the superposition. At this point we have produced a new $\ket{\mu_i}$ state, which encodes the data of the users $n-1$,\ldots,$i+1$ and $i$.

The additional dual-rail CNOT gate we have not commented on yet is essential for a correct procedure. It will make sure that no entanglement with the original qubit is left. With it, the evolution 
\begin{equation}
\alpha_i\ket{\mathit{0}}\ket{v}\ket{\mu_{j}}\ket{v}+\beta_i\ket{v}\ket{\mathit{0}}\ket{v}\ket{\mu_{j}} \stackrel{\small{CNOT}}{\longrightarrow} \alpha_i\ket{\mathit{0}}\ket{v}\ket{\mu_{j}}\ket{v}+\beta_i\ket{\mathit{0}}\ket{v}\ket{v}\ket{\mu_{j}},
\end{equation}
guarantees that the $i$th photon can be found with certainty in the upper qubit port (the qubit state is always $\ket{\mathbf{0}}$). A photodetector can confirm everything went as desired. If this last CNOT step were not taken, the state would not be transferred to the qudit, but shared with the original qubit. That could spoil later operations at the receiver because of unwanted distinguishability. 

After all the multiplexing blocks, the OAM qudit, originally in $\ket{\mathit{0}}$, will encode in its winding number the digit represented by the $n$ qubits. The $i$th block adds to the OAM qudit the corresponding power of two of the binary expansion of $\ell$.

If there are superpositions, each binary number has its own probability amplitude. In the final state, each possible qubit combination is represented by an $\ket{\ell}$ state with the corresponding probability amplitude. The transmitted qudit will be 
\begin{equation}
\sum_{\ell=0}^{2^n-1} \hspace{2ex}\prod_{k=0}^{n-1}\left( \alpha^{b_{\ell}^k\oplus 1}_k \beta^{\hspace{1ex} b_{\ell}^k}_k \right)\ket{\ell},
\end{equation}
where $\ell=b_{\ell}^{n-1}\cdots b_{\ell}^{k} \cdots b_{\ell}^{0}$ in binary and $\oplus$ represents modulo 2 addition. The procedure is valid for arbitrary inputs and preserves any previous entanglement between channels. Due to the Dove parity check that lies at the heart of the procedure, the incorporation has to be made from the most to the least significant qubits. 

\subsection{Demultiplexing}
Similarly, it is possible to build the inverse block, either from the realization that all the elements are, again, reversible, or by undoing all the steps of the transmitter one by one, extracting one qubit at a time. Reversibility is not ruined by the detection steps. They are always on known states that can be exactly reproduced at the other side of the communication.

\begin{figure}[h]
\centering
\includegraphics{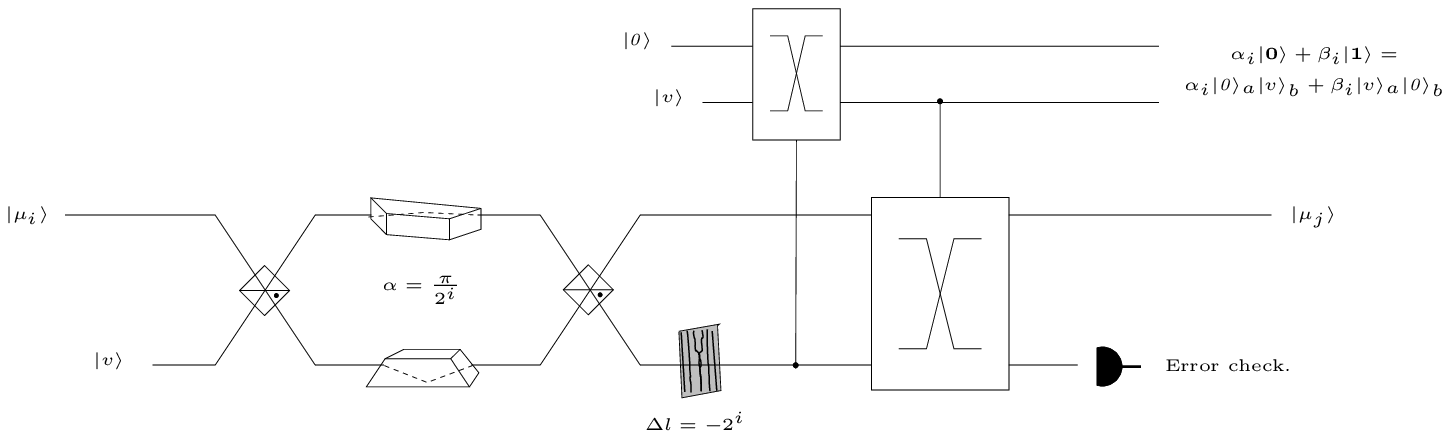}
\caption{OAM demultiplexing block for the $i$th qubit.\label{OAMDEMUX}} 
\end{figure}

Figure \ref{OAMDEMUX} presents the resulting demultiplexer circuit (OAM DEMUX). All the superposed states enter the upper port. Only the terms carrying the $+2^i$ indicator of the $\ket{\mathbf{1}}$ state of the $i$th qubit will suffer a $\pi$ phase shift that will change their paths. The part of the photon in the lower port still carries the information of the rest of the qubits, so the qubit cannot be extracted just by a path separation. The sequence of the two dual-rail CNOTs transfers the state to the upper qubit. The first CNOT modifies the path of the new photon or not depending on the content of the $i$th qubit. The second CNOT completes the transfer by erasing the $i$th qubit from the OAM qudit.

The hologram will subtract the $2^i$ contribution so that at the next step the OAM qudit can be processed by a different Dove prism with a higher K. After the second dual-rail CNOT, the qudits from the same states that were separated as a result of the incorporation of the $i$th qubit, interfere again, restoring their previous value. The extraction must mirror the encoding and suffers from the same limitation on the operation order. The recovery must start from the least significant qubit. At each stage, the most significant qubits of the remaining superposition are separated from the extracted, less significant, qubit. This is again a consequence of the Dove prisms operation.

\begin{figure}[h]
\centering
\includegraphics{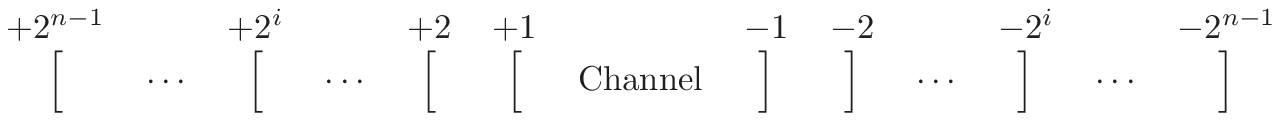}
\caption{Nested increasing level (OAM number) MUX and DEMUX operation .\label{nested}} 
\end{figure}

Figure \ref{nested} shows the nested configuration necessary for a correct operation. The lack of efficient general OAM sorters imposes a particular block order and a sequential recovery at the receiver.

An additional user of index $n$ can always be added with an extra $+2^n$ stage before the existing scheme and a $-2^n$ stage after it. There is no limit to the maximum number of users, but they cannot be incorporated into the scheme once the multiplexing has started. The least significant qubits can be reserved to allocate unexpected new users up to a limit. They need not to be occupied for a correct operation and this reservation will add flexibility. Nevertheless, practical restrictions will appear, making the use of high $\ell$ OAM states inadvisable and there will be a trade-off between flexible and easy to realize systems.

The first and the last blocks can be avoided if the photon of the first qubit is used as the qudit carrier. The path qubit can be converted into OAM dual-rail qubit with the converter circuit of Figure \ref{converter} (without the last hologram) with the same resulting qudit state as the corresponding OAM MUX block. We only loose the possibility of an error check that the erasure of the original qubit gave. Similarly, the last qubit is already an OAM dual-rail qubit. It can be easily converted into a path dual-rail qubit either by the inverse of the converter circuit or by the corresponding OAM DEMUX block. The only difference is the additional pair of CNOT gates that provides a state transfer of the remaining qubit into a new photon. The qudit that transmitted the information must then be in $\ket{\mathit{0}}$. It can be measured to check for operation errors. However, it is likely that the CNOT gates will be the limiting factor for the scheme's implementation. Reducing their number is probably the safest option. 

We can see that, from the total $4n$ CNOT gates of the scheme ($2$ gates for each of the $n$ blocks, both at the transmitter and the receiver), 4 can be saved if the the transmitted photon is recycled from the $n$th qubit, $\ket{\psi_{n-1}}$. For the most likely applications, with just a few qubits, this can be an important saving. 

\section{Other applications}
\label{apps} 
We have seen how the OAM MUX and DEMUX circuits can increase the information transmission capacity of a quantum communication system by sending data from more than one source over the same link. Apart from this advantage in data rate and user coordination, their capability to convert between different quantum information encodings can be useful in other situations. In this Section, we discuss the applications of the OAM multiplexing setups to two different problems: OAM sorting and the construction of quantum arithmetic units. 

\subsection{Dual rail CNOT gates and efficient sorting}
An OAM sorter is an optical system that can separate $\ket{\ell}$ states attending to their winding number. There are different proposals, but no universal efficient separation exists. A good review on OAM sorting can be found in \cite{LPB02}, which proposes the ingenious sorter that is behind the main blocks of our model. This sorter allows to identify arbitrary OAM states by chaining $n$ parity separation stages in an exponentially branching setup. The different values of $\ell$ are separated attending to their remainder in divisions by different powers of two.

There is an unexpected side result from the MUX and DEMUX scheme. The OAM DEMUX can act as an efficient OAM sorter with a gate complexity that grows logarithmically with the number of states to be sorted. In our OAM MUX, there are $m=2^n$ possible OAM values and the number of stages grows linearly with the number of qubits, $n$. A general sorter is straightforward to derive from the same considerations. 

If we have an unknown input state $\ket{\ell}$, where it is guaranteed that $\ell\leq M$, and feed it into a DEMUX, the resulting qubits will encode the value of $\ell$. Furthermore, superpositions will be kept. If the qubits are measured, the operation will correspond to the regular projective use of OAM sorters, which detect only one value of $\ell$ and act as a measurement. For this case, we will need a DEMUX with $\lceil \log_2{M} \rceil$ blocks, where $\lceil x \rceil$ represents the ceiling function, which rounds the number $x$ up to the immediately higher integer.

The inclusion of optical dual-rail CNOT gates allows to avoid an exponential branching. The blocks apply the corresponding $\Delta\ell$ corrections so that the same interferometer can act on what in the previous scheme were two different branches. The $n$ resulting qubits hold all the necessary information about the winding number, $\ell$, of the input.

Unfortunately, efficient optical CNOT gates have not been built yet. We can also propose an OAM sorter with quantum non-demolition measurements instead of CNOT gates. Optical quantum non-demolition, QND, measurements able to tell the existence of a photon without destroying it, although also difficult to realize, seem closer than CNOT gates. This kind of QND measurement would determine a path for the photon. The superposition between states with different parity would be destroyed, but a photon in the port can still be in a superposition of the different OAM states that take that path.

In most OAM sorting, it is not required to keep superpositions. We are only interested in a projective measurement of the OAM state. The transfer to a series of qubits that was an important part of the DEMUX operation is not strictly necessary for the sorter. If we could ascertain the presence or absence of a photon in a particular path mode, we would be able to determine the parity of its winding number. In that case, we can replace the first CNOT gate and the qubit of the DEMUX by a classical bit storing the result of the QND measurement (0 if the photon is in the upper port, 1 if it is in the lower port). Instead of the second CNOT gate, we can use a classical switch that changes the photon paths if the bit is in 1. The $\Delta\ell$ correction, like in the DEMUX, prepares the state for the next sorting stage. Each block is equivalent to measuring the value of one of the qubits. At the end, the state is projected into a single OAM value, which can be determined by reading the bits with the result of the individual QND measurements. This scheme also has $\lceil \log_2{M} \rceil$ stages.

It is worth to note that, conversely, a more general sorter could improve the flexibility of the OAM MUX and the OAM combiner. If we had at our disposal a sorter that could separate OAM states according to the quotient of the division of their winding number by a given power of two, $\ell\hspace{1ex} div\hspace{1ex} 2^i$, the restriction in the user order at the transmitter and the receiver could be lifted. 

All these details suggest that the CNOT operation, QND measurement and efficient OAM sorting have a similar power for quantum information processing. Optical CNOT gates provide direct QND measurement and, conversely, QND measurements can be part of efficient schemes to provide optical CNOT gates \cite{NM04}. Given the difficulties in finding both efficient optical CNOT gates and QND measurement schemes, this will probably mean that realizing an efficient OAM sorter that does not require exponential resources will be far from trivial. 

\subsection{Quantum arithmetic}
The basic units of OAM multiple access and the structure of its circuits can be modified to provide efficient quantum arithmetic in higher dimensional spaces, followed by a later conversion to the multiple qubits space. The number of gates and simplicity of those setups might constitute them as a viable alternative for elementary quantum operations. 

If the channels of the OAM multiplexer are not independent users but the different qubits of the binary representation of a number, different arithmetic operations can be performed on the OAM states. The multiplexing stage will convert them into a more manipulable domain. In the demultiplexing stage, the result of the operation will be recovered and put into the corresponding quantum registers. We have already anticipated this use by speaking of most and least significant qubits instead of speaking of channels.  

All the operations suggested in this section will assume non-negative integer operands that are encoded into the corresponding $\ket{\ell}$ state, where $\ell$ is the decimal representation of the qubit sequence. The operands will all have the same size of $n$ qubits. 

\subsubsection{Adder}
\begin{figure}[h]
\centering
\includegraphics{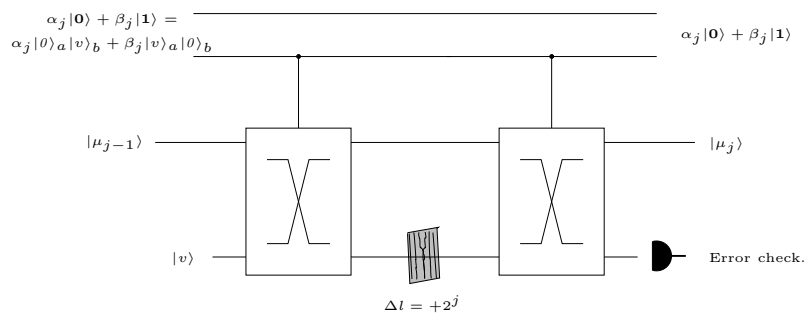}
\caption{Adder block.\label{adder}} 
\end{figure}

After the multiplexing of the $n$ qubits, the output qudit will be a superposition of the integers encoded by each qubit combination. For a non-negative integer, $N$, those states can be selectively added the different powers of two that define the binary representation of a second number $M$. Both $N$ and $M$ can be in a superposition of different values. A sequence of $n$ blocks like the one in Figure \ref{adder} will add to the original state the appropriate power of two for each of the terms of the superposition of $M$ and, at the output, the corresponding superposition of sum states will appear. 

Here, the qudit is only directed to the hologram when the $j$th qubit is $\ket{\mathbf{1}}$ and only in this case will the winding number be increased. For this application, once the first number is in an OAM state, no particular block order is required.

Notice that the second operand is not erased or completely absorbed into the OAM qudit. This is, in fact, a requisite for reversible operation. A quantum adder must keep one of the operands to be able to revert the operation. 

The new OAM state can later be decoded by an OAM demultiplexer. The only provision to be made is to consider that the new value of the sum can be as high as $2^n-1+2^n-1=2^{n+1}-2$, so that an additional qubit must be extracted with an extra block for $i=n$ after the usual $i=n-1$ step. Alternatively, we could impose the most significant qubits of the inputs to be $\ket{\mathbf{0}}$.

The complexity of the adder circuit is of $n$ MUX blocks, $n+1$ DEMUX blocks and $n$ adders. In terms of CNOT gates, the most critical element, this amounts to $6n+2$ gates, or $6n-2$ gates if the first qubit of the MUX and last qubit of the DEMUX are encoded in the same photon that carried the information and suffered the addition. 

This circuit can be compared to the full adders of previous proposals such as \cite{VBE96}. Both adder circuits scale linearly with the input size. The main advantage of our scheme is that carries need not to be explicitly taken into account, allowing for conceptually simpler circuits. 

\subsubsection{OAM ALUs}
The Arithmetic Logic Unit, ALU, is one of the basic building blocks in the standard Von Neumann computer architecture. The ALU performs all the logical operations that transform the data. Among its tasks, the ALU must perform integer arithmetic operations such as addition and multiplication. 

With the provided OAM MUX and DEMUX architectures, all the operations can be translated to the integer OAM domain. In the previous section, an OAM adder has been described. A multiplier unit can also be built from a similar model. If the hologram of Figure \ref{adder} is replaced by a $\times2^j$ block of operation 
\begin{equation}
\ket{\ell}\stackrel{\times2^j}{\longrightarrow}\ket{2^j\ell},
\end{equation}
the final state will encode the integer value $N\times M$. Then, a DEMUX can recover the product into qubits. 

If the multiplying block could be performed by an optical element, it would constitute an important complexity saving when compared with the usual $O(n^2)$ blocks of the usual multiplication algorithms \cite{VBE96}. However, to the best of the authors' knowledge, no element has been experimentally demonstrated to provide efficient multiplication of arbitrary OAM photon states while preserving the rest of the photon's parameters. Strong nonlinearities and conversion of multiple photons are possible candidates for such an interaction, but seem unlikely to be efficient at the single photon level. 

OAM adders and multipliers based on OAM multiplexing would have the advantage of using the same custom blocks. The use of the same elements for different functions is important when it comes to scaling and production. If efficient multiplexers are built, basic arithmetic operations could be efficiently performed in the OAM domain. 

\section{Outlook}
\label{outlook}
OAM can be used to implement two different kinds of optical free-space multiplexers. The first proposal, the OAM combiner, puts together $n$ photons into the same path. The second one, the OAM MUX, takes all the information from $n$ qubits into a single photon with the help of optical CNOT gates. 

Both OAM setups can have their own range of application. The OAM combiner is already possible to build and has to its favour that separate qubits are less vulnerable to errors. Qubit losses or decoherence happen to the individual qubits. The OAM MUX needs of more advanced technology that includes an optical CNOT gate. The main advantage of this setup is its ability to carry, in theory, arbitrary amounts of information in one single photon. This is a mixed blessing. By having all the information in a single photon, if it is absorbed, all the qubits are lost.

In both cases there is an important limitation: recovery must be sequential, in a certain prescribed order. This restriction is a direct consequence of the OAM sorting procedure we use. More flexible OAM sorters would permit more powerful OAM multiplexing schemes. 

Although, in theory, both proposals could be used for any number of qubits, holograms and the other optical elements are likely to allow only for a limited maximum number of $\ket{\ell}$ states to be created and processed. For clarity, we have used $\ell$ from 0 to $2^n$ in the OAM MUX. However, easier to manipulate $\ell$ values from $-2^{n-1}$ to $+2^{n-1}$, similar to those of the OAM combiner, could be used for the scheme. Apart from that, lossy elements can be avoided. There exist techniques used to design more efficient OAM sorters \cite{WXL03} that could be translated to our scheme. 

All the schemes could also be used for classical optical multiplexing, where losses are not so important and a part of the light can be divided into the different users so that parallel extraction is easier to perform. It should be possible to perform experiments for a few users. Classical OAM transmission systems have already used with good results values of OAM in the range of $\pm16\hbar$ \cite{GCP04} and there are quantum schemes that could produce different superpositions of photons with OAM up to $\pm10\hbar$ or even higher \cite{MTT01}. This means that three or four channels could be sent together using a qubit merger with the present technology. 

\acknowledgements{}
We would like to thank Graham S. McDonald for helpful insights into the effects of beamsplitters and Dove prisms on the OAM of light. J.C. Garc\'ia Escart\'in would also like to thank Clara I. Osorio for an interesting discussion on the limitations of the optical equipment normally used in OAM experiments. This research has been funded by project TEC2007-67429-C02-01 of the Spanish MICINN.

\newcommand{\noopsort}[1]{} \newcommand{\printfirst}[2]{#1}
  \newcommand{\singleletter}[1]{#1} \newcommand{\switchargs}[2]{#2#1}


\begin{thebibliography}{27}
\expandafter\ifx\csname natexlab\endcsname\relax\def\natexlab#1{#1}\fi
\expandafter\ifx\csname bibnamefont\endcsname\relax
  \def\bibnamefont#1{#1}\fi
\expandafter\ifx\csname bibfnamefont\endcsname\relax
  \def\bibfnamefont#1{#1}\fi
\expandafter\ifx\csname citenamefont\endcsname\relax
  \def\citenamefont#1{#1}\fi
\expandafter\ifx\csname url\endcsname\relax
  \def\url#1{\texttt{#1}}\fi
\expandafter\ifx\csname urlprefix\endcsname\relax\def\urlprefix{URL }\fi
\providecommand{\bibinfo}[2]{#2}
\providecommand{\eprint}[2][]{\url{#2}}

\bibitem[{\citenamefont{Poppe et~al.}(2004)\citenamefont{Poppe, Fedrizzi,
  Ursin, B\"{o}hm, L\"{o}runser, Maurhardt, Peev, Suda, Kurtsiefer, Weinfurter
  et~al.}}]{PFU04}
\bibinfo{author}{\bibfnamefont{A.}~\bibnamefont{Poppe}},
  \bibinfo{author}{\bibfnamefont{A.}~\bibnamefont{Fedrizzi}},
  \bibinfo{author}{\bibfnamefont{R.}~\bibnamefont{Ursin}},
  \bibinfo{author}{\bibfnamefont{H.}~\bibnamefont{B\"{o}hm}},
  \bibinfo{author}{\bibfnamefont{T.}~\bibnamefont{L\"{o}runser}},
  \bibinfo{author}{\bibfnamefont{O.}~\bibnamefont{Maurhardt}},
  \bibinfo{author}{\bibfnamefont{M.}~\bibnamefont{Peev}},
  \bibinfo{author}{\bibfnamefont{M.}~\bibnamefont{Suda}},
  \bibinfo{author}{\bibfnamefont{C.}~\bibnamefont{Kurtsiefer}},
  \bibinfo{author}{\bibfnamefont{H.}~\bibnamefont{Weinfurter}},
  \bibnamefont{et~al.}, \bibinfo{journal}{Optics Express}
  \textbf{\bibinfo{volume}{12}}, \bibinfo{pages}{3865} (\bibinfo{year}{2004}).

\bibitem[{\citenamefont{Landry et~al.}(2007)\citenamefont{Landry, van
  Houwelingen, Beveratos, Zbinden, and Gisin}}]{LHB07}
\bibinfo{author}{\bibfnamefont{O.}~\bibnamefont{Landry}},
  \bibinfo{author}{\bibfnamefont{J.~A.~W.} \bibnamefont{van Houwelingen}},
  \bibinfo{author}{\bibfnamefont{A.}~\bibnamefont{Beveratos}},
  \bibinfo{author}{\bibfnamefont{H.}~\bibnamefont{Zbinden}}, \bibnamefont{and}
  \bibinfo{author}{\bibfnamefont{N.}~\bibnamefont{Gisin}}, \bibinfo{journal}{Journal of the
  Optical Society of America B} \textbf{\bibinfo{volume}{24}}, \bibinfo{pages}{398}
  (\bibinfo{year}{2007}).

\bibitem[{\citenamefont{Villoresi et~al.}(2008)\citenamefont{Villoresi,
  Jennewein, Tamburini, Aspelmeyer, Bonato, Ursin, Pernechele, Luceri, Bianco,
  Zeilinger et~al.}}]{VJT08}
\bibinfo{author}{\bibfnamefont{P.}~\bibnamefont{Villoresi}},
  \bibinfo{author}{\bibfnamefont{T.}~\bibnamefont{Jennewein}},
  \bibinfo{author}{\bibfnamefont{F.}~\bibnamefont{Tamburini}},
  \bibinfo{author}{\bibfnamefont{M.}~\bibnamefont{Aspelmeyer}},
  \bibinfo{author}{\bibfnamefont{C.}~\bibnamefont{Bonato}},
  \bibinfo{author}{\bibfnamefont{R.}~\bibnamefont{Ursin}},
  \bibinfo{author}{\bibfnamefont{C.}~\bibnamefont{Pernechele}},
  \bibinfo{author}{\bibfnamefont{V.}~\bibnamefont{Luceri}},
  \bibinfo{author}{\bibfnamefont{G.}~\bibnamefont{Bianco}},
  \bibinfo{author}{\bibfnamefont{A.}~\bibnamefont{Zeilinger}},
  \bibnamefont{et~al.}, \bibinfo{journal}{New Journal of Physics}
  \textbf{\bibinfo{volume}{10}}, \bibinfo{pages}{033038}
  (\bibinfo{year}{2008}).

\bibitem[{\citenamefont{Sklar}(1983)}]{Skl83}
\bibinfo{author}{\bibfnamefont{B.}~\bibnamefont{Sklar}},
  \bibinfo{journal}{IEEE Communications Magazine}
  \textbf{\bibinfo{volume}{21}}, \bibinfo{pages}{6} (\bibinfo{year}{1983}).

\bibitem[{\citenamefont{Sklar}(2001)}]{Skl01}
\bibinfo{author}{\bibfnamefont{B.}~\bibnamefont{Sklar}},
  \emph{\bibinfo{title}{Digital Communications}} (\bibinfo{publisher}{Prentice
  Hall}, \bibinfo{address}{Upper Saddle River, New Jersey 07458},
  \bibinfo{year}{2001}), \bibinfo{edition}{2nd} ed.

\bibitem[{\citenamefont{Brassard et~al.}(2003)\citenamefont{Brassard,
  Bussieres, Godbout, and Lacroix}}]{BBG03}
\bibinfo{author}{\bibfnamefont{G.}~\bibnamefont{Brassard}},
  \bibinfo{author}{\bibfnamefont{F.}~\bibnamefont{Bussieres}},
  \bibinfo{author}{\bibfnamefont{N.}~\bibnamefont{Godbout}}, \bibnamefont{and}
  \bibinfo{author}{\bibfnamefont{S.}~\bibnamefont{Lacroix}},
  \bibinfo{journal}{Proceedings of the SPIE Applications of Photonic Technology
  6} \textbf{\bibinfo{volume}{5260}}, \bibinfo{pages}{149}
  (\bibinfo{year}{2003}).

\bibitem[{\citenamefont{Ortigosa-Blanch and Capmany}(2006)}]{OC06}
\bibinfo{author}{\bibfnamefont{A.}~\bibnamefont{Ortigosa-Blanch}}
  \bibnamefont{and} \bibinfo{author}{\bibfnamefont{J.}~\bibnamefont{Capmany}},
  \bibinfo{journal}{Physical Review A} \textbf{\bibinfo{volume}{73}},
  \bibinfo{eid}{024305} (\bibinfo{year}{2006}).

\bibitem[{\citenamefont{{Gibson} et~al.}(2004)\citenamefont{{Gibson},
  {Courtial}, {Padgett}, {Vasnetsov}, {Pas'ko}, {Barnett}, and
  {Franke-Arnold}}}]{GCP04}
\bibinfo{author}{\bibfnamefont{G.}~\bibnamefont{{Gibson}}},
  \bibinfo{author}{\bibfnamefont{J.}~\bibnamefont{{Courtial}}},
  \bibinfo{author}{\bibfnamefont{M.~J.} \bibnamefont{{Padgett}}},
  \bibinfo{author}{\bibfnamefont{M.}~\bibnamefont{{Vasnetsov}}},
  \bibinfo{author}{\bibfnamefont{V.}~\bibnamefont{{Pas'ko}}},
  \bibinfo{author}{\bibfnamefont{S.~M.} \bibnamefont{{Barnett}}},
  \bibnamefont{and}
  \bibinfo{author}{\bibfnamefont{S.}~\bibnamefont{{Franke-Arnold}}},
  \bibinfo{journal}{Optics Express} \textbf{\bibinfo{volume}{12}},
  \bibinfo{pages}{5448} (\bibinfo{year}{2004}).

\bibitem[{\citenamefont{Allen et~al.}(2003)\citenamefont{Allen, Barnett, and
  Padgett}}]{ABP03}
\bibinfo{author}{\bibfnamefont{L.}~\bibnamefont{Allen}},
  \bibinfo{author}{\bibfnamefont{S.}~\bibnamefont{Barnett}}, \bibnamefont{and}
  \bibinfo{author}{\bibfnamefont{M.}~\bibnamefont{Padgett}},
  \emph{\bibinfo{title}{{Optical Angular Momentum}}}
  (\bibinfo{publisher}{Institute of Physics Publishing, Bristol, UK},
  \bibinfo{year}{2003}).

\bibitem[{\citenamefont{{Molina-Terriza}
  et~al.}(2007)\citenamefont{{Molina-Terriza}, {Torres}, and {Torner}}}]{MTT07}
\bibinfo{author}{\bibfnamefont{G.}~\bibnamefont{{Molina-Terriza}}},
  \bibinfo{author}{\bibfnamefont{J.~P.} \bibnamefont{{Torres}}},
  \bibnamefont{and} \bibinfo{author}{\bibfnamefont{L.}~\bibnamefont{{Torner}}},
  \bibinfo{journal}{Nature Physics} \textbf{\bibinfo{volume}{3}},
  \bibinfo{pages}{305} (\bibinfo{year}{2007}).

\bibitem[{\citenamefont{G\"{o}tte et~al.}(2008)\citenamefont{G\"{o}tte,
  O'Holleran, Preece, Flossmann, Franke-Arnold, Barnett, and Padgett}}]{GOP08}
\bibinfo{author}{\bibfnamefont{J.~B.} \bibnamefont{G\"{o}tte}},
  \bibinfo{author}{\bibfnamefont{K.}~\bibnamefont{O'Holleran}},
  \bibinfo{author}{\bibfnamefont{D.}~\bibnamefont{Preece}},
  \bibinfo{author}{\bibfnamefont{F.}~\bibnamefont{Flossmann}},
  \bibinfo{author}{\bibfnamefont{S.}~\bibnamefont{Franke-Arnold}},
  \bibinfo{author}{\bibfnamefont{S.~M.} \bibnamefont{Barnett}},
  \bibnamefont{and} \bibinfo{author}{\bibfnamefont{M.~J.}
  \bibnamefont{Padgett}}, \bibinfo{journal}{Optics Express}
  \textbf{\bibinfo{volume}{16}}, \bibinfo{pages}{993} (\bibinfo{year}{2008}).

\bibitem[{\citenamefont{{Turnbull} et~al.}(1996)\citenamefont{{Turnbull},
  {Robertson}, {Smith}, {Allen}, and {Padgett}}}]{TRS96}
\bibinfo{author}{\bibfnamefont{G.~A.} \bibnamefont{{Turnbull}}},
  \bibinfo{author}{\bibfnamefont{D.~A.} \bibnamefont{{Robertson}}},
  \bibinfo{author}{\bibfnamefont{G.~M.} \bibnamefont{{Smith}}},
  \bibinfo{author}{\bibfnamefont{L.}~\bibnamefont{{Allen}}}, \bibnamefont{and}
  \bibinfo{author}{\bibfnamefont{M.~J.} \bibnamefont{{Padgett}}},
  \bibinfo{journal}{Optics Communications} \textbf{\bibinfo{volume}{127}},
  \bibinfo{pages}{183} (\bibinfo{year}{1996}).

\bibitem[{\citenamefont{Heckenberg et~al.}(1992)\citenamefont{Heckenberg,
  McDuff, Smith, and White}}]{HMS92}
\bibinfo{author}{\bibfnamefont{N.~R.} \bibnamefont{Heckenberg}},
  \bibinfo{author}{\bibfnamefont{R.}~\bibnamefont{McDuff}},
  \bibinfo{author}{\bibfnamefont{C.~P.} \bibnamefont{Smith}}, \bibnamefont{and}
  \bibinfo{author}{\bibfnamefont{A.~G.} \bibnamefont{White}},
  \bibinfo{journal}{Optics Letters} \textbf{\bibinfo{volume}{17}},
  \bibinfo{pages}{221} (\bibinfo{year}{1992}).

\bibitem[{\citenamefont{{Abramochkin} and {Volostnikov}}(1991)}]{AV91}
\bibinfo{author}{\bibfnamefont{E.}~\bibnamefont{{Abramochkin}}}
  \bibnamefont{and}
  \bibinfo{author}{\bibfnamefont{V.}~\bibnamefont{{Volostnikov}}},
  \bibinfo{journal}{Optics Communications} \textbf{\bibinfo{volume}{83}},
  \bibinfo{pages}{123} (\bibinfo{year}{1991}).

\bibitem[{\citenamefont{{Mair} et~al.}(2001)\citenamefont{{Mair}, {Vaziri},
  {Weihs}, and {Zeilinger}}}]{MVW01}
\bibinfo{author}{\bibfnamefont{A.}~\bibnamefont{{Mair}}},
  \bibinfo{author}{\bibfnamefont{A.}~\bibnamefont{{Vaziri}}},
  \bibinfo{author}{\bibfnamefont{G.}~\bibnamefont{{Weihs}}}, \bibnamefont{and}
  \bibinfo{author}{\bibfnamefont{A.}~\bibnamefont{{Zeilinger}}},
  \bibinfo{journal}{Nature} \textbf{\bibinfo{volume}{412}},
  \bibinfo{pages}{313} (\bibinfo{year}{2001}).

\bibitem[{\citenamefont{Molina-Terriza
  et~al.}(2001)\citenamefont{Molina-Terriza, Torres, and Torner}}]{MTT01}
\bibinfo{author}{\bibfnamefont{G.}~\bibnamefont{Molina-Terriza}},
  \bibinfo{author}{\bibfnamefont{J.~P.} \bibnamefont{Torres}},
  \bibnamefont{and} \bibinfo{author}{\bibfnamefont{L.}~\bibnamefont{Torner}},
  \bibinfo{journal}{Physical Review Letters} \textbf{\bibinfo{volume}{88}},
  \bibinfo{pages}{013601} (\bibinfo{year}{2001}).

\bibitem[{\citenamefont{Leach et~al.}(2002)\citenamefont{Leach, Padgett,
  Barnett, Franke-Arnold, and Courtial}}]{LPB02}
\bibinfo{author}{\bibfnamefont{J.}~\bibnamefont{Leach}},
  \bibinfo{author}{\bibfnamefont{M.~J.} \bibnamefont{Padgett}},
  \bibinfo{author}{\bibfnamefont{S.~M.} \bibnamefont{Barnett}},
  \bibinfo{author}{\bibfnamefont{S.}~\bibnamefont{Franke-Arnold}},
  \bibnamefont{and} \bibinfo{author}{\bibfnamefont{J.}~\bibnamefont{Courtial}},
  \bibinfo{journal}{Physical Review Letters} \textbf{\bibinfo{volume}{88}},
  \bibinfo{pages}{257901} (\bibinfo{year}{2002}).

\bibitem[{\citenamefont{Arlt et~al.}(1998)\citenamefont{Arlt, Dholakia, Allen,
  and Padgett}}]{ADA98}
\bibinfo{author}{\bibfnamefont{J.}~\bibnamefont{Arlt}},
  \bibinfo{author}{\bibfnamefont{K.}~\bibnamefont{Dholakia}},
  \bibinfo{author}{\bibfnamefont{L.}~\bibnamefont{Allen}}, \bibnamefont{and}
  \bibinfo{author}{\bibfnamefont{M.~J.} \bibnamefont{Padgett}},
  \bibinfo{journal}{Journal of Modern Optics} \textbf{\bibinfo{volume}{45}},
  \bibinfo{pages}{1231} (\bibinfo{year}{1998}).

\bibitem[{\citenamefont{Alipasha~Vaziri and Zeilinger}(2002)}]{VWZ02}
\bibinfo{author}{\bibfnamefont{A.} \bibnamefont{Vaziri}}
  \bibinfo{author}{\bibfnamefont{G.}~\bibnamefont{Weihs}},
  \bibnamefont{and}
  \bibinfo{author}{\bibfnamefont{A.}~\bibnamefont{Zeilinger}},
  \bibinfo{journal}{Journal of Optics B: Quantum and Semiclassical Optics}
  \textbf{\bibinfo{volume}{4}}, \bibinfo{pages}{S47} (\bibinfo{year}{2002}).

\bibitem[{\citenamefont{Yao et~al.}(2006)\citenamefont{Yao, Franke-Arnold,
  Courtial, Padgett, and Barnett}}]{YFC06}
\bibinfo{author}{\bibfnamefont{E.}~\bibnamefont{Yao}},
  \bibinfo{author}{\bibfnamefont{S.}~\bibnamefont{Franke-Arnold}},
  \bibinfo{author}{\bibfnamefont{J.}~\bibnamefont{Courtial}},
  \bibinfo{author}{\bibfnamefont{M.~J.} \bibnamefont{Padgett}},
  \bibnamefont{and} \bibinfo{author}{\bibfnamefont{S.~M.}
  \bibnamefont{Barnett}}, \bibinfo{journal}{Optics Express}
  \textbf{\bibinfo{volume}{14}}, \bibinfo{pages}{13089} (\bibinfo{year}{2006}).

\bibitem[{\citenamefont{St\"{u}tz et~al.}(2007)\citenamefont{St\"{u}tz,
  Gr\"{o}blacher, Jennewein, and Zeilinger}}]{SGJ07}
\bibinfo{author}{\bibfnamefont{M.}~\bibnamefont{St\"{u}tz}},
  \bibinfo{author}{\bibfnamefont{S.}~\bibnamefont{Gr\"{o}blacher}},
  \bibinfo{author}{\bibfnamefont{T.}~\bibnamefont{Jennewein}},
  \bibnamefont{and}
  \bibinfo{author}{\bibfnamefont{A.}~\bibnamefont{Zeilinger}},
  \bibinfo{journal}{Applied Physics Letters} \textbf{\bibinfo{volume}{90}},
  \bibinfo{eid}{261114} (\bibinfo{year}{2007}).

\bibitem[{\citenamefont{Wang et~al.}(2005)\citenamefont{Wang, Sun, Shum, and
  Yin}}]{WSS05}
\bibinfo{author}{\bibfnamefont{Q.}~\bibnamefont{Wang}},
  \bibinfo{author}{\bibfnamefont{X.}~\bibnamefont{Sun}},
  \bibinfo{author}{\bibfnamefont{P.}~\bibnamefont{Shum}}, \bibnamefont{and}
  \bibinfo{author}{\bibfnamefont{X.~J.} \bibnamefont{Yin}},
  \bibinfo{journal}{Optics Express} \textbf{\bibinfo{volume}{13}},
  \bibinfo{pages}{10285} (\bibinfo{year}{2005}).

\bibitem[{\citenamefont{Born and Wolf}(1997)}]{BW97}
\bibinfo{author}{\bibfnamefont{M.}~\bibnamefont{Born}} \bibnamefont{and}
  \bibinfo{author}{\bibfnamefont{E.}~\bibnamefont{Wolf}},
  \emph{\bibinfo{title}{Principles of Optics}} (\bibinfo{publisher}{Cambridge
  University Press, Cambridge, UK}, \bibinfo{year}{1997}),
  \bibinfo{edition}{sixth} ed.

\bibitem[{\citenamefont{Gonz\'{a}lez et~al.}(2006)\citenamefont{Gonz\'{a}lez,
  Molina-Terriza, and Torres}}]{GMT06}
\bibinfo{author}{\bibfnamefont{N.}~\bibnamefont{Gonz\'{a}lez}},
  \bibinfo{author}{\bibfnamefont{G.}~\bibnamefont{Molina-Terriza}},
  \bibnamefont{and} \bibinfo{author}{\bibfnamefont{J.~P.}
  \bibnamefont{Torres}}, \bibinfo{journal}{Optics Express}
  \textbf{\bibinfo{volume}{14}}, \bibinfo{pages}{9093} (\bibinfo{year}{2006}).

\bibitem[{\citenamefont{Xue et~al.}(2001)\citenamefont{Xue, Wei, and
  Kirk}}]{XWK01}
\bibinfo{author}{\bibfnamefont{X.}~\bibnamefont{Xue}},
  \bibinfo{author}{\bibfnamefont{H.}~\bibnamefont{Wei}}, \bibnamefont{and}
  \bibinfo{author}{\bibfnamefont{A.~G.} \bibnamefont{Kirk}},
  \bibinfo{journal}{Optics Letters} \textbf{\bibinfo{volume}{26}},
  \bibinfo{pages}{1746} (\bibinfo{year}{2001}).

\bibitem[{\citenamefont{Padgett and Allen}(2002)}]{PA02}
\bibinfo{author}{\bibfnamefont{M.~J.} \bibnamefont{Padgett}} \bibnamefont{and}
  \bibinfo{author}{\bibfnamefont{L.}~\bibnamefont{Allen}},
  \bibinfo{journal}{Journal of Optics B: Quantum and Semiclassical Optics}
  \textbf{\bibinfo{volume}{4}}, \bibinfo{pages}{S17} (\bibinfo{year}{2002}).

\bibitem[{\citenamefont{Vedral et~al.}(1996)\citenamefont{Vedral, Barenco, and
  Ekert}}]{VBE96}
\bibinfo{author}{\bibfnamefont{V.}~\bibnamefont{Vedral}},
  \bibinfo{author}{\bibfnamefont{A.}~\bibnamefont{Barenco}}, \bibnamefont{and}
  \bibinfo{author}{\bibfnamefont{A.}~\bibnamefont{Ekert}},
  \bibinfo{journal}{Physical Review A} \textbf{\bibinfo{volume}{54}},
  \bibinfo{pages}{147} (\bibinfo{year}{1996}).

\bibitem[{\citenamefont{Nemoto et~al.}(2004)\citenamefont{Nemoto, and Munro}}]{NM04}
\bibinfo{author}{\bibfnamefont{K.}~\bibnamefont{Nemoto}},\bibnamefont{and}
  \bibinfo{author}{\bibfnamefont{W.J.}~\bibnamefont{Munro}},
  \bibinfo{journal}{Physical Review Letters} \textbf{\bibinfo{volume}{93}},
  \bibinfo{pages}{250502} (\bibinfo{year}{2004}).

\bibitem[{\citenamefont{Wei et~al.}(2003)\citenamefont{Wei, Xue, Leach,
  Padgett, Barnett, Franke-Arnold, Yao, and Courtial}}]{WXL03}
\bibinfo{author}{\bibfnamefont{H.}~\bibnamefont{Wei}},
  \bibinfo{author}{\bibfnamefont{X.}~\bibnamefont{Xue}},
  \bibinfo{author}{\bibfnamefont{J.}~\bibnamefont{Leach}},
  \bibinfo{author}{\bibfnamefont{M.}~\bibnamefont{Padgett}},
  \bibinfo{author}{\bibfnamefont{S.}~\bibnamefont{Barnett}},
  \bibinfo{author}{\bibfnamefont{S.}~\bibnamefont{Franke-Arnold}},
  \bibinfo{author}{\bibfnamefont{E.}~\bibnamefont{Yao}}, \bibnamefont{and}
  \bibinfo{author}{\bibfnamefont{J.}~\bibnamefont{Courtial}},
  \bibinfo{journal}{Optics Communications} \textbf{\bibinfo{volume}{223}},
  \bibinfo{pages}{117} (\bibinfo{year}{2003}).

\end{thebibliography}
\end{document}